\documentclass[journal=apchd5,manuscript=article]{achemso}

\usepackage{chemformula} 
\usepackage[T1]{fontenc} 
\usepackage{fancyref}
\usepackage{hyperref}
\usepackage[normalem]{ulem}

\setcitestyle{round}
\setcitestyle{numbers,sort,nocompress}

\newcommand{\onlinecite}[1]{\hspace{-1 ex} \nocite{#1}\citenum{#1}}

\author{Matthew Proctor}
\affiliation[Imperial College London]{Department of Mathematics, Imperial College London, London, SW7 2AZ, UK}
\alsoaffiliation{Department of Physics, Imperial College London, London, SW7 2AZ, UK}
\email{matthew.proctor12@imperial.ac.uk}

\author{Richard V. Craster}
\affiliation[Imperial College London]{Department of Mathematics, Imperial College London, London, SW7 2AZ, UK}

\author{Stefan A. Maier}
\affiliation[Munich]{Chair in Hybrid Nanosystems, Nanoinstitut M{\"u}nchen, Faculty of Physics, Ludwig-Maximilians-Universit{\"a}t M{\"u}nchen, 80539, M{\"u}nchen, Germany}
\alsoaffiliation{Department of Physics, Imperial College London, London, SW7 2AZ, UK}

\author{Vincenzo Giannini}
\affiliation[Madrid]{Instituto de Estructura de la Materia (IEM), Consejo Superior de Investigaciones Cient\'{i}ficas (CSIC), Serrano 121, 28006, Madrid, Spain}

\author{Paloma A. Huidobro}
\affiliation[Lisbon]{Instituto de Telecomunica{\c c}{\~o}es, Insituto Superior Tecnico-University of Lisbon, Avenida Rovisco Pais 1,1049-001 Lisboa, Portugal}
\alsoaffiliation{Department of Physics, Imperial College London, London, SW7 2AZ, UK}

\title[Exciting Pseudospin Dependent Edge States in Plasmonic Metasurfaces]
  {Exciting Pseudospin Dependent Edge States in Plasmonic Metasurfaces}

\abbreviations{1D, 2D, SSH, LSP, QSA, CDA, PTI}

\begin{document}

\begin{abstract}
 We study a plasmonic metasurface that supports pseudospin dependent edge states confined at a subwavelength scale, considering full electrodynamic interactions including retardation and radiative effects. The spatial symmetry of the lattice of plasmonic nanoparticles gives rise to edge states with properties reminiscent of the quantum spin Hall effect in topological insulators. However, unlike the spin-momentum locking characteristic of topological insulators, these modes are not purely unidirectional and their propagation properties can be understood by analysing the spin angular momentum of the electromagnetic field, which is inhomogenous in the plane of the lattice. The local sign of the spin angular momentum determines the propagation direction of the mode under a near-field excitation source. We also study the optical response under far-field excitation and discuss in detail the effects of radiation and retardation. 
\end{abstract}

Topological insulators are materials which are insulating in the bulk but which have conduction surface states protected against disorder \cite{hasan2010colloquium}. The remarkable properties of these states in electronic systems has inspired the search for photonic topological insulators (PTIs), which aim to guide and manipulate photons with the same level of control and efficiency \cite{ozawa2019topological, rider2019perspective, khanikaev2017two,lu2014topological,silveirinha2019proof}. Systems which possess these effects whilst preserving time reversal symmetry are appealing as they do not require complicated experimental setups such as strong magnetic fields or bianisotropic coupling. Motivated by this, a proposal to emulate the quantum spin Hall (QSH) effect in photonic crystals was presented by Wu and Hu in Ref. \onlinecite{wu2015scheme}. Effects reminiscent of the QSH phase such as a band inversion between dipolar and quadrupolar modes, and pseudospin dependent edge states are realised but, rather than relying on the time reversal symmetric pairs characteristic of electronic systems, they instead rely on the spatial symmetry of the lattice structure. As a result, the edge states have a reduction in backscattering over trivial ones \cite{orazbayev2018chiral, orazbayev2019quantitative}. The method has since been applied to a variety of bosonic systems \cite{jiang2019manipulation,chaunsali2018subwavelength, barik2016two,yves2017crystalline,yves2017topological}, and has recently experimentally been demonstrated in the visible regime \cite{peng2019probing}. 

The combination of topological effects with plasmonics offers the possibility of precisely controlling light on the the nanoscale. The strong enhancement and localisation of electric fields due to localised surface plasmon (LSP) resonances \cite{maier2007plasmonics} is a widely employed platform for light confinement on the nanoscale \cite{giannini2011plasmonic, kravets2018plasmonic}. Plasmonic metasurfaces can be formed by arranging plasmonic nanoparticles in two-dimensional (2D) lattices, where the LSPs become delocalised across the whole metasurface as collective resonances. The optical properties of metasurfaces are then determined by the individual nanoparticle elements as well as the geometry of the lattice \cite{meinzer2014plasmonic,baur2018hybridization}. The tunable optical properties of metasurfaces makes them versatile tools for the manipulation of light on the nanoscale \cite{wang2018rich,monticone2017metamaterial}. For instance, appropriately designed plasmonic metasurfaces can host spin dependent directional states which can couple to valley excitons when interfaced to 2D materials \cite{chervy2018room}.

One-dimensional (1D) chains of dielectric and metallic nanoparticles were some of the first systems used for the investigation of topological phases in nanophotonics \cite{poddubny2014topological,slobozhanyuk2015subwavelength,downing2017topological, downing2018topological, kruk2017edge, kruk2019nonlinear, wu2019dynamic}, in particular systems analogous to the Su-Schrieffer-Heeger (SSH) model, which hosts topologically protected edge states in 1D. In the plasmonic chain, initial studies into these topological states were limited to the quasistatic approximation (QSA) \cite{ling2015topological} despite the well known radiative effects which are of great importance for large enough nanoparticles and retardation which is important at large lattice periods \cite{weber2004propagation, koenderink2006complex} or at very small periods, where higher order multipolar effects occur \cite{park2004surface}. Indeed, beyond the quasistatic limit, the plasmonic SSH chain becomes non-Hermitian and band structures are distorted compared to the quasistatic model, with effects such as polariton splitting at the light line \cite{ pocock2018topological}. In addition, the ubiquitous bulk-edge correspondence of topological insulators has been shown to break down due to retardation \cite{pocock2019bulk}. 2D plasmonic systems, including graphene and arrays of plasmonic nanoparticles have also been considered for hosting topological states. Time reversal broken effects reliant on magnetic fields have been proposed for graphene plasmons \cite{pan2017topologically,jin2017infrared}. Honeycomb lattices of plasmonic nanoparticles have been investigated in the quasistatic limit, where a direct analogy between the tight binding model in graphene and nearest neighbour approximation in plasmonics can be made \cite{weick2013dirac, han2009dirac, wang2016existence}. More recently, theoretical and experimental investigations have shown how the long range, retarded interactions affect the physical behaviour of the system \cite{fernique2019plasmons,guo2019lasing}. Here, we consider the lattice geometry proposed in Ref. \onlinecite{wu2015scheme}, shown in \autoref{fig:lattice_layout}, to study pseudospin dependent edge states in plasmonic metasurfaces. This scheme has been considered in 2D arrays of plasmonic nanoparticles in the QSA \cite{honari2019topological}. In this work, we show the importance of going beyond this approximation and study the realisation of electromagnetic modes resembling the QSH effect on a plasmonic metasurface, including full electrodynamic interactions in our description of the system.

We use semi-analytical techniques to investigate plasmonic metasurfaces. These consist of a triangular lattice with unit cells containing six nanoparticles arranged in a hexagon, \autoref{fig:lattice_layout}a. We begin by outlining the coupled dipole method used to model these arrays and then we investigate the behaviour of the modes supported by the metasurface. By calculating the eigenmodes of the infinite lattice, we characterise the band inversion process which occurs between a \textit{shrunken} phase (where nanoparticles in the unit cell are displaced towards the centre) and an \textit{expanded} phase (where they are displaced outwards from the centre). The response of the infinite system is also studied under far-field excitation.

We then consider the interface between two phases in a semi-infinite ribbon layout to elucidate the nature of the edge states that emerge due to the band inversion between the two phases. We characterise their pseudospin dependence, showing that these modes are more appropriately characterized by means of the spin angular momentum of the electromagnetic fields, and probe these states in the near-field with a magnetic dipole source. By investigating the effect of the source position, we unambiguously show that, unlike the unidirectional edge states characteristic of topological insulators, the directionality of these edge states depends on the source position. Finally, we highlight the radiative and retardation effects on the edge states.

\section*{Design and set up of the plasmonic metasurface}\label{sec:design}
\begin{figure}
    \includegraphics[width=\textwidth]{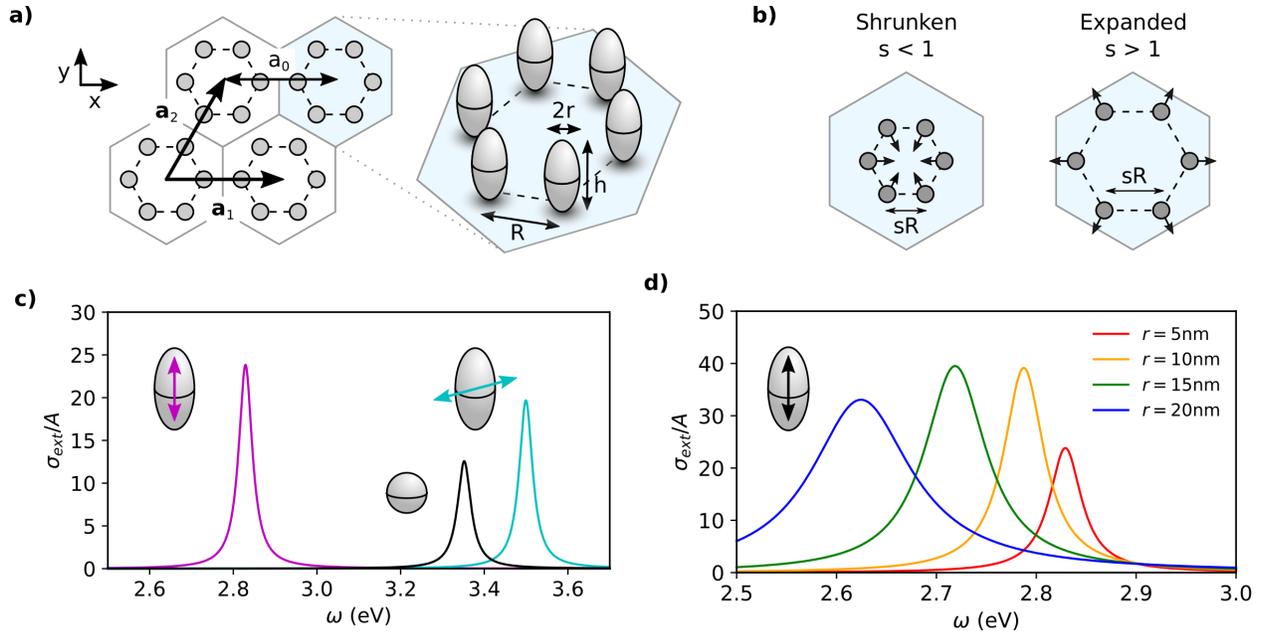}
    \caption{Metal rods are arranged on a plane to form a plasmonic metasurface, and their optical response is modelled including radiative and retarded effects.  a) Layout of the lattice of plasmonic nanoparticles, including a close up of the arrangement of particles in the unit cell, b) The perturbation of the unit cell into shrunken and expanded phases with scaling parameter $s$. $s=1$ corresponds to the unperturbed honeycomb lattice. c) Extinction cross sections $\sigma_{ext}$, normalized to geometrical cross section, of silver spheroidal nanoparticles showing the splitting of in-plane (blue) and out-of-plane (pink) resonances, d) $\sigma_{ext}$ of nanoparticles with increasing radius, showing the radiative broadening and redshifting of the plasmon resonance.}
    \label{fig:lattice_layout}
\end{figure}

The metasurface we consider here consists of a 2D array of metal nanorods (modelled as spheroidal nanoparticles) arranged in the lattice shown in \autoref{fig:lattice_layout}a. The unit cell contains six nanoparticles of radius $r$ and height $h$ arranged in a hexagon separated by nearest neighbour spacing $R$ with lattice vectors $\textbf{a}_1$ and $\textbf{a}_2$, as shown in the figure. In the regime $R > 3r$, the nanoparticles can be considered as point dipoles and higher order resonances can be neglected \cite{maier2003optical,weber2004propagation}. A nanoparticle at position $\textbf{d}_i$ with polarisability $\alpha(\omega)$ supports a dipole moment $\textbf{p}_i$ when excited by an external field  $\textbf{E}_i$. For a lattice of nanoparticles, a coupled dipole equation can be written which describes the dipole moment of a nanoparticle induced by an external field plus the sum of all neighbouring dipole moments,
\begin{align} \label{eq:CDA}
    \frac{1}{\alpha(\omega)}\textbf{p}_i = \textbf{E}_i + \sum_{i \neq j} \hat{\textbf{G}}(\textbf{d}_i - \textbf{d}_j, \omega) \textbf{p}_j,
\end{align}
where the dyadic Green's function $\hat{\textbf{G}}$, describes the interactions between point dipoles, and is given by, 
\begin{align}
    \hat{\textbf{G}}(\textbf{d}_i - \textbf{d}_j, \omega) = k^2\frac{e^{ikd}}{d} \left[
    \left(
    1 + \frac{i}{kd} - \frac{1}{k^2d^2}
    \right)\hat{\textbf{I}}
    - 
    \left(
    1 + \frac{3i}{kd} - \frac{3}{k^2d^2}
    \right)\textbf{n}\otimes\textbf{n}
    \right].
\end{align}
Here, $k=\sqrt{\epsilon_B}\omega/c$ is the wavenumber of the surrounding medium, with $\epsilon_B$ the permittivity of the surrounding environment (which is assumed to be the vacuum throughout this work, $\epsilon_B=1$), and  $d = |\textbf{d}_i - \textbf{d}_j|$ is the distance between nanoparticles, with $\textbf{n} = (\textbf{d}_i - \textbf{d}_j)/|\textbf{d}_i - \textbf{d}_j|$ the unit vector in the direction along the line that joins two nanoparticles.

In equation \ref{eq:CDA}, the polarisability, $\alpha(\omega)$, describes the optical response of an individual nanoparticle. The static polarisability for a spheroidal nanoparticle is written, 
\begin{align}
    \alpha_{s}(\omega) = \frac{V}{4\pi}\frac{\epsilon(\omega) - 1}{1 + L\left(\epsilon(\omega) - 1\right)},
\end{align}
where $\epsilon(\omega)$ is the dielectric function of the metal, $V$ is the spheroid volume, and $L$ is the static geometrical factor which is dependent on the radius and height of the nanoparticle; for a sphere $L=\frac{1}{3}$  \cite{moroz2009depolarization}. The dielectric function of the nanoparticles is given by the Drude model,
\begin{align}
    \epsilon(\omega) = \epsilon_\infty - \frac{\omega_p^2}{\omega^2 + i \omega\gamma}.
\end{align}
Throughout the work we consider silver nanoparticles, with $\epsilon_\infty = 5$, $\omega_p = 8.9$~eV and $\gamma = 1/17$~fs $\approx 0.039$~eV \cite{yang2015optical}. The static polarisability neglects radiative effects which are essential for describing larger nanoparticles.  We take this into account by means of the modified long wavelength approximation (MLWA), 
\begin{align}
    \alpha_{\text{MLWA}}(\omega) = \dfrac{\alpha_s(\omega)}{1 - \dfrac{k^2}{l_E}D\alpha_s(\omega) - i\dfrac{2k^3}{3}\alpha_s(\omega)},
\end{align}

where $l_E$ is the spheroid major axis half length and $D$ is a dynamic geometrical factor; $D=1$ for a sphere \cite{moroz2009depolarization}. The importance of the radiative correction for spheroidal silver nanoparticles is exemplified in \autoref{fig:lattice_layout}c. The extinction cross section $\sigma_{\text{ext}}$ for a spheroidal nanoparticle with radius $r = 5$~nm and height $h=20$~nm, as well as a spherical nanoparticle with radius $r=5$~nm is shown. We normalise to the cross sectional area $A$ perpendicular to the dipole moment. The nanoparticle supports two resonance modes: one where the dipole is aligned with the minor axis (in-plane) and one with the major axis (out-of-plane). The out-of-plane resonance becomes redshifted and well separated in frequency from the in-plane resonance, which allows us to investigate the out-of-plane and in-plane modes separately. The increased radiative effect on a single nanoparticle is demonstrated by the extinction cross section for larger radii, up to $r=20$~nm, where the resonance becomes broader and continues to be redshifted, \autoref{fig:lattice_layout}d. Whilst the static polarisabilty adequately describes the behaviour of the smallest nanoparticle sizes, the MLWA incorporates the effects of dynamic depolarisation and the radiative correction, and is necessary to correctly model particles of radius above $\sim10$ nm.

\section*{Spectral response of the metasurface}\label{sec:spectral}

We start by considering the optical response of the plasmonic metasurface in the expanded and shrunken phases, \autoref{fig:lattice_layout}b. We do so by setting up an infinite lattice of nanoparticles and applying periodic boundary conditions to \autoref{eq:CDA} by writing the external electric field $\mathbf{E}$, and dipole moments $\mathbf{p}$, as periodic Bloch functions. The following system of equations can then be written,
\begin{align}
\left(\frac{1}{\alpha(\omega)} - \hat{\textbf{H}}(\textbf{k}, \omega)\right)\cdot\textbf{p} = \textbf{E}, \label{eqn:eigenvalue}
\end{align}
where the interaction matrix $\hat{\textbf{H}}(\textbf{k}, \omega)$ has elements,
\begin{align}
    H_{ij}=
    \begin{cases}
    \sum\limits_{\textbf{R}} \hat{\textbf{G}}(\textbf{d}_i - \textbf{d}_j + \textbf{R}, \omega) \hspace{2px} e^{i\textbf{k}\cdot\textbf{R}} & i \neq j\\
    \sum\limits_{|\textbf{R}|\neq0} \hat{\textbf{G}}(\textbf{R}, \omega) \hspace{2px} e^{i\textbf{k}\cdot\textbf{R}} & i = j
    \end{cases},
    \label{eqn:hamiltonian}
\end{align}
with $\textbf{q}$, the Bloch wavevector and $\mathbf{R}$, the lattice site positions. We note that the interaction matrix is a $6\times6$ matrix since we restrict our study to the out-of-plane modes, such that there is a single degree of freedom for each particle in the unit cell. The in-plane modes of a honeycomb lattice of plasmonic nanoparticles have been studied elsewhere \cite{wang2016existence} and being well shifted in frequencies, they are completely decoupled from the out-of-plane modes. Finally, we note that sums in the interaction matrix are conditionally convergent due to the slowly decaying $1/r$ term and so additional manipulation is required to converge these expressions (see Methods). 

\begin{figure}
    \includegraphics[width=0.43\textwidth]{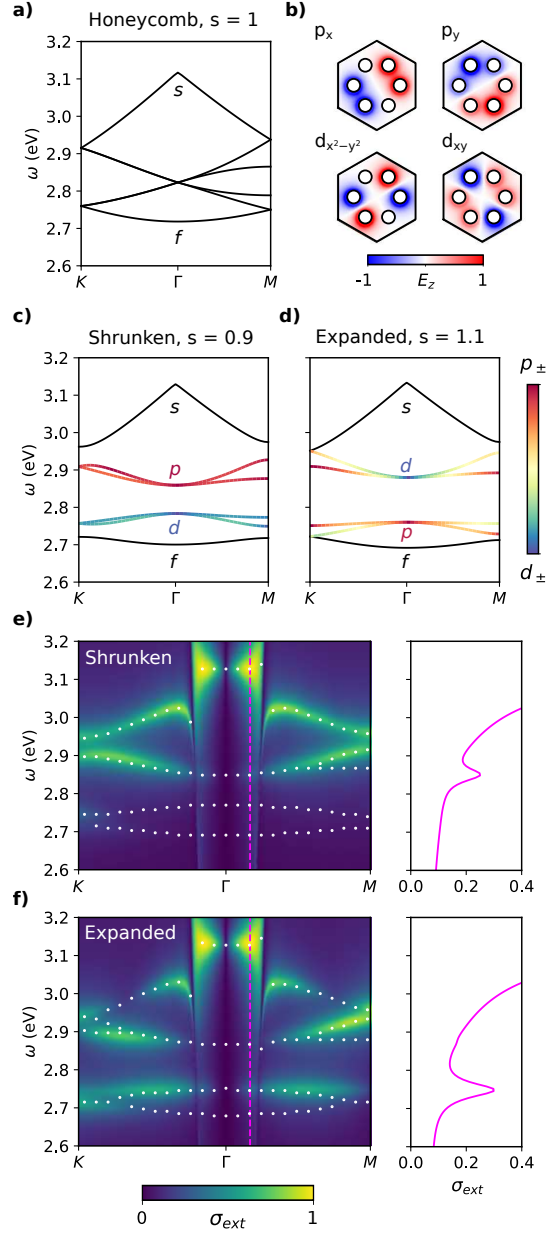}
    \caption{Exciting plasmonic metasurfaces from the far-field. Nanoparticles have radius $r = 5$~nm and height $h = 20$~nm, and the lattice constant $a_0 = 60$~nm. a) Bulk dispersion relations of the out-of-plane modes of the honeycomb plasmonic lattice, including all neighbours in the quasistatic approximation (QSA), b) Normalised, out-of-plane electric fields of the quadrupolar and dipolar modes of the metasurface (at lower and higher energies, respectively), including full GF interactions. c) Dispersion relations for the shrunken ($s=0.9$) and d) expanded ($s=1.1$) systems. The bands are coloured according to their dipolar ($p$) or quadrupolar ($d$) nature, showing the band inversion at $\Gamma$ between shrunken and expanded phases. The monopolar ($s$) and hexapolar ($f$) bands are not involved in the inversion. e), f) Extinction cross sections, normalised to the maximum, under excitation with an external field including full retarded Green's function (GF) interactions and radiative polarisabilities for the shrunken and expanded phases. The dots highlight all the modes in the system, some of which are dark in an external field. $\sigma_{\text{ext}}$ at a fixed wavevector (purple dotted line) is shown in the right hand plots.}
    \label{fig:far_field}
\end{figure}

\autoref{fig:far_field} shows the spectral response of the plasmonic metasurface under study. We consider nanoparticles with radius $r = 5$~nm and height $h = 20$~nm, nearest neighbour spacing $R = 20$~nm and lattice constant $a_0 = 60$~nm. To calculate the eigenvalues and eigenvectors of the periodic lattice we solve \autoref{eqn:eigenvalue} without an incident field. Initially, in order to keep the eigenvalue problem linear we take the QSA and only consider the quickly decaying $1/d^3$ term in the Green's function. However, we choose to go beyond the nearest neighbour approximation and include interactions between all particles in the lattice \cite{zhen2008collective}. The band structure of a metasurface with the nanoparticles arranged in a honeycomb lattice is shown in \autoref{fig:far_field}a. Instead of using the conventional rhombic unit cell, we take the larger hexagonal cell such that the Brillouin zone (BZ) of the honeycomb lattice becomes folded and the Dirac points at $K$ and $K'$ are mapped onto each other to create a doubly degenerate point at $\Gamma$ \cite{wu2015scheme}, as shown in \autoref{fig:far_field}a. Whereas the original honeycomb lattice is formed of two triangular sublattices, the system is now formed of six sublattices corresponding to the six nanoparticles in the unit cell; meaning there are six bands present in the band structure. In the QSA, assuming nearest neighbour interactions, the Dirac points of the honeycomb plasmonic lattice occur at the surface plasmon resonance frequency $w_{sp}$ \cite{wang2016existence} and the band structure is symmetrical about this frequency \cite{honari2019topological}. This is no longer the case due to the sublattice symmetry breaking interaction term between particles of the same sublattice in neighbouring unit cells. \autoref{fig:far_field}c shows the band structures for metasurfaces with shrunken (left) and expanded (right) unit cells. The bands are labelled as $s,\, p,\, d,\, f$, indicating monopolar, dipolar, quadrupolar and hexapolar characters. This was determined by calculating the overlap with the eigenstates of an isolated hexagon of nanoparticles. The dipole and quadrupole modes of the lattice are shown in \autoref{fig:far_field}b. In both the shrunken and expanded phases, we see how the the ordering of the modes is opposite to that obtained in photonic crystals and other bosonic analogues \cite{wu2015scheme, gorlach2018far, yves2017crystalline, yves2017topological}, with the monopolar mode being the highest in energy and the hexapolar being the lowest in energy. This is due to the different electromagnetic properties of the two systems:  while in the photonic crystal the permittivity is positive and constant,  in the plasmonic
system the permittivity is negative and dispersive. For the metasurface considered here, the
bonding or antibonding character of the coupling between the plasmonic nanoparticles determines the mode ordering, as we explain in more detail later. Next, we show that near the centre of the BZ, there is a mode inversion between the shrunken and expanded phases. This is evident from the colour scale, which encodes the dipolar/quadrupolar character of the modes. In the shrunken phase ($s<1$), the band above the band gap is dipolar and the one below is quadrupolar (\autoref{fig:far_field}c). On the other hand, for the expanded phase ($s > 1$) they become inverted at $\Gamma$ (\autoref{fig:far_field}d).  
The degeneracy of the bands above and below the gap at $\Gamma$ for the shrunken and expanded lattices suggests linear combinations of these modes can be taken, $p_\pm = (p_x \pm ip_y)$ and $d_\pm =  (d_{x^2 - y^2} \pm id_{xy})$, which correspond to pseudospins; taking positive or negative combinations gives clockwise or anticlockwise rotations \cite{wu2015scheme}. 

After characterising the eigenstates in the QSA, we now consider the full electrodynamic interaction between the nanoparticles by including all terms in the Green's function, and explore retardation and radiative effects by calculating the extinction cross section of the system when excited by an external field. We show the response of the system in the shrunken and expanded phase over the BZ in \autoref{fig:far_field}e-f (contour plots). The incident plane wave is defined as $\textbf{E}=(E_x, E_y, E_z)$, where the field components satisfy Maxwell's equations and $k_0^2 = k_\parallel^2 + k_z^2$. Above the light line, the wave is propagating in the $z$ direction, $k_z = \sqrt{k_0^2 - k_\parallel^2}$ and below the light line it is evanescent, $k_z = i\sqrt{k_\parallel^2 - k_0^2}$, which allows us to probe the eigenstates in this region of the spectrum. Starting at $\Gamma$, the incident field has no $z$-component meaning out-of-plane dipoles cannot be excited and the extinction cross section is zero. As soon as we move away from $\Gamma$, we begin to excite out-of-plane polarised modes. The highest energy mode corresponds to a monopolar mode, with all dipole moments in the unit cell pointing in the same direction. The energy ordering can be understood from the ordering of bonding and anti-bonding modes in a plasmonic dimer. The bonding mode lies at a higher energy compared to the anti-bonding mode, as has been shown theoretically \cite{nordlander2004plasmon} and experimentally \cite{funston2009plasmon}. This is a consequence of the  relative orientation of the dipoles: in the antibonding mode the dipoles are antiparallel which minimizes the radiated electric field  and hence the energy, whereas in the bonding mode the dipoles are parallel and hence sustain a larger electric field and appear at higher energies. When plasmonic nanoparticles are arranged in a lattice with more than one particle per unit cell, this ordering is mantained, as has also been shown elsewhere \cite{fernique2019plasmons, mann2018manipulating}. Notably, this is not an effect of retardation and will arise in any near-field coupled ensemble of plasmonic nanoparticles, as shown by our quasistatic results (\autoref{fig:far_field}c,d) and in Refs. \cite{nordlander2004plasmon,fernique2019plasmons, mann2018manipulating}.
 
In the infinite lattice, the bonding nature of the monopolar mode across the whole system causes it to be highly radiant and so dominates the response of the lattice for propagating waves. Nevertheless, we are still able to examine the peaks in extinction cross section at lower frequencies. By symmetry arguments, only the dipolar mode is excitable by a plane wave. The extinction cross section of the system agrees with the characterisation of the modes in the QSA. In the shrunken phase, the higher energy dipolar mode is visible in the extinction cross section whereas in the expanded phase a band inversion occurs. This inversion between dipolar and quadrupolar modes close to the BZ centre constitutes a signature to distinguish both metasurface phases, since it can be detected experimentally by far-field measurements \cite{gorlach2018far}. In the right panels of \autoref{fig:far_field}e-f we show the extinction spectrum at a fixed incident momentum, $\textbf{k}$: the resonance peak corresponding to the dipolar mode is visible for the shrunken structure at higher energies than for the expanded structure.  In \autoref{fig:far_field}e-f, we plot the loci of peaks in the spectral function based on the effective polarisability (see Methods), to make all of the modes visible, even those which are dark in an external field. These peaks qualitatively agree with the QSA dispersion relation with only a slight redshift due to the radiative correction given that all the length scales in the system are very subwavelength. Retardation effects are most apparent in the highest energy band. The coupling of the plasmonic mode with free photons causes the strong polariton-like splitting at the light line \cite{koenderink2006complex, zhen2008collective}. Since the monopolar (bonding) mode is strongly radiant, it shows the largest interaction with the light line. We also observe how the radiative broadening of the mode grows larger going from $\Gamma$ towards the light line; this has been observed in 1D plasmonic chains and the effect is much greater in this 2D lattice due to constructive interference between dipole moments in the monopolar mode \cite{fernique2019plasmons}.
The polariton splitting and radiative broadening importantly do not affect the band inversion at $\Gamma$.  Finally, the lowest energy band is a hexapolar mode where all dipole moments are in anti-phase meaning such a mode cannot be excited by a plane wave, and it is not visible in the extinction cross section for most of the BZ.

\section*{Pseudospin edge states in finite systems}\label{sec:pseudospin}

\begin{figure}
    \includegraphics[width=0.5\textwidth]{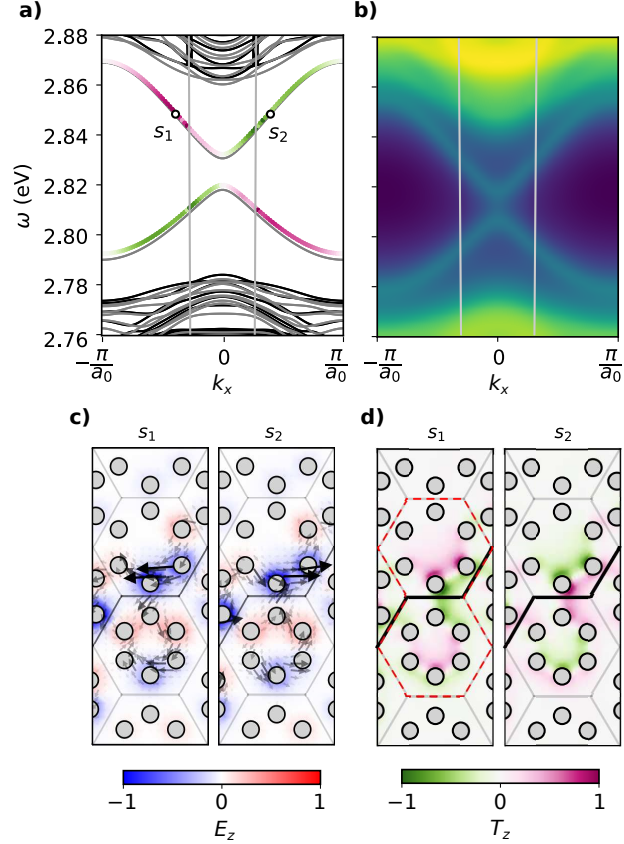}
    \caption{Pseudospin modes along the interface between metasurfaces in expanded and shrunken phases. a) Dispersion relations of a semi-infinite ribbon with a 28 unit cell region in the expanded phase ($s=1.1$) cladded with two 6 unit cell regions in the shrunken phase ($s=0.9$). Results obtained in the QSA including all neighbours are shown in grey and results including full GF interactions are shown in black, with both in the non-radiative regime. We highlight the edge modes in this case with colours according to their pseudospin, characterised by spin angular momentum $T$. b) Spectral function of the retarded, radiative system with Drude losses $\gamma=0.01$~eV. The frequency shift in the edge states compared to the QSA in a) is due to the radiative effects, c) Normalised out-of-plane electric field $E_z$ for $s_1$ and $s_2$, in the metasurface plane. The arrows show the time averaged Poynting vector $\textbf{S}$, demonstrating the directionality of the edge states, d) Normalised spin angular momentum $T$ for modes $s_1$ and $s_2$, marked in a). $T$ is integrated in the red region colour the edge states in a)}
    \label{fig:ribbon}
\end{figure}

After characterising the modes in the metasurface, we now look at the edge states between regions of different phases. As discussed earlier, the band inversion between different phases is reminiscent of the QSH effect and \sout{thus pseudospin dependent, }directional edge states are anticipated. In order to study the edge states, we consider an interface between the two different phases of the metasurface. In the calculations, a finite ribbon of a region in the expanded phase is cladded with regions in the shrunken phase so that we have a finite structure along the direction $\textbf{a}_2$, and we apply Bloch periodicity along $\textbf{a}_1$. \autoref{fig:ribbon}a shows the band structure of the ribbon, calculated with retardation (black lines), and also in the QSA (grey lines). In the case of full electrodynamics interactions, the Green's function was linearised by letting $\omega = \omega_{sp}$ and the static polarisability was assumed (see Methods). Two edge states can be seen clearly within the band gap (we colour the retarded solution, as we will explain below). Importantly, the edge states do not join the bulk modes, rather they reconnect with each other at the edges of the BZ. This is a fundamental distinction with topological edge states since it means it is possible, by a continuous change of parameters, to remove these states from the band gap; which indicates that these edge states are not resistant to backscattering \cite{qian2018topology}. In \autoref{fig:ribbon}b, we also plot the spectral function of the system with radiative corrections in the polarisability, which accounts for the frequency shift (in this plot we let the Drude losses $\gamma=0.01$~eV to improve the visibility of the edge states). Similarly to the bulk dipolar and quadrupolar bands, we see the edge states do not interact strongly with the light line although the radiative broadening and redshift is apparent. Since the pseudospin effect is reliant on the $C_{6v}$ symmetry of the bulk system which is necessarily broken at the interface between the shrunken and expanded region there will always be a `minigap' between the edge states at $\Gamma$ which will never fully close. The amount of perturbation between shrunken and expanded phase determines the size of the bulk band gap and as well as the size of the minigap. To minimise the gap, the perturbation along the edge can be graded; the edge states are then excitable across the whole band gap \cite{honari2019topological, chaunsali2018subwavelength}.

To investigate the nature of the edge modesand to elucidate their excitation under point sources we now look at the two eigenstates of the expanded/shrunken interface with opposite group velocity, and at a frequency $\omega$ in the upper band. These are shown as points $s_1$ and $s_2$ in \autoref{fig:ribbon}a. We first plot the time averaged Poynting vector $\textbf{S} = \frac{1}{2}\text{Re}(\textbf{E}\times\textbf{H}^*)$ along with the normalised out-of-plane electric field $E_z$, in the plane of the metasurface $z=0$, in \autoref{fig:ribbon}c. This demonstrates that the edge modes are confined to the interface between the two regions. The Poynting vector (arrows) characterises the flow of electromagnetic energy, which is in opposite directions for wavevectors with opposite sign. In inhomogeneous, dispersive media it can be argued that the Poynting vector does not necessarily accurately characterise a spin or pseudospin \cite{bliokh2017opticala}. More appropriately, we also consider the spin angular momentum in the plane of the lattice, $\textbf{T} = \text{Im}(\textbf{E}^*\times\textbf{E} + \textbf{H}^*\times\textbf{H})$ \cite{bliokh2017optical}. Since the dipole moments are aligned out of the plane, at $z=0$ the field components are $E_z, H_x, H_y$ and the spin angular momentum is then $T_z = \text{Im}(\textbf{H}^*\times\textbf{H})$. This spin angular momentum quantifies the degree of elliptical polarization of the magnetic field and its handedness, similar to the studies for photonic crystals in Refs. \cite{oh2018chiral,deng2017transverse}. We plot the normalised spin over the same region as the Poynting vector in \autoref{fig:ribbon}c, where its inhomogenous character across the lattice is evident, with varying magnitude and sign across the interface. To determine the pseudospin dependence of the edge states, we integrate the spin angular momentum $T$ over the region shown in red in \autoref{fig:ribbon}d, for wavevectors $k_\parallel$ across the whole BZ, and we plot the edge states with this colour code in the dispersion relation in \autoref{fig:ribbon}a. At the edges of the BZ there is maximum mixing between pseudospins and the integral of $T = 0$, but as we move towards the centre we see either edge state acquires an opposite pseudospin. This pattern of opposite pseudospins travelling in opposite directions is akin to the QSH effect. Again, we note that since in this system the two edge modes are linked through a pseudo-time reversal operator \cite{wu2015scheme}, which is only rigourously defined at $\Gamma$, there is not complete protection against backscattering. 

\begin{figure}
    \includegraphics[width=0.9\textwidth]{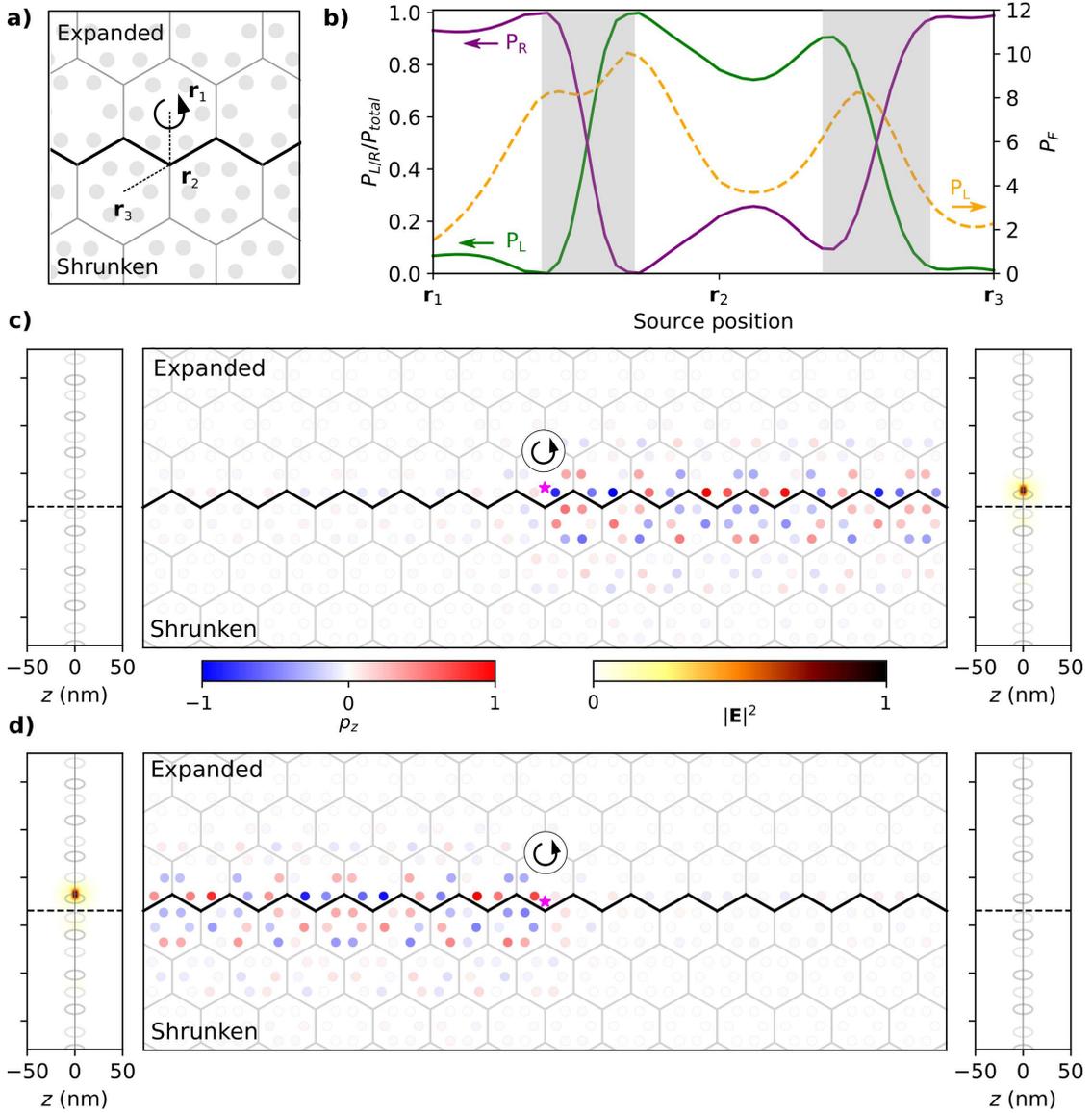}
    \caption{Exciting pseudospin edge states with near-field probes. a) A right circularly polarised magnetic dipole is placed at positions along the path shown to excite the edge state between expanded and shrunken regions in the metasurface, b) The fraction of power travelling left and right, $P_{L/R}$, as well as the Purcell factor $P_F$ are dependent on source position across the edge. Although the source is right circularly polarised, it will excite an edge state in the opposite direction for some positions along the edge. (Regions where the coupled dipole approximation does not hold are shaded), c) Pseudospin edge state excited by a source at the optimal position for a right travelling mode. The middle panel shows the dipole moments in the plane of the metasurface ($xy$ plane). The left and right panels show the normalised electric field intensity $|\textbf{E}|^2$ perpendicular to the metasurface ($yz$ plane), showing the mode is strongly confined in the out-of-plane direction $z$.  d) A source with the same polarisation as in c) is placed at the least optimal position showing excitation in the opposite direction.}
    \label{fig:directionality}

\end{figure}

We now consider the excitation of edge modes with a localised source, and its relation to the spin angular momentum. 
The sign of the spin angular momentum $T$ is positive in the region within the hexagon of nanoparticles immediately above the edge for the mode $s_1$, as shown by the purple area in \autoref{fig:ribbon}d, left panel. However, at a point directly on the edge, shown by the black line in \autoref{fig:ribbon}d, the sign of $T$ switches.
Importantly, this means that the local handedness of the elliptical polarization of the magnetic field for each of the edge modes is inhomogeneous in the plane of the array, as expected for a confined mode in a complex environment. This position dependence of the spin angular momentum has implications on the excitation of unidirectional modes by point sources, which we show by modelling a finite array of nanoparticles with an interface between an expanded and shrunken region. To fully understand how the propagation along the edge is dependent on the source and its position, we consider an interface with zero material losses; the edge state is prevented from reflecting off the hard boundary with the vacuum by slowly increasing material losses at these boundaries. We place a right circularly polarised magnetic dipole source with magnetic field $\textbf{H} = H_x + iH_y$ at the various positions shown in \autoref{fig:directionality}a. We choose the excitation frequency from \autoref{fig:ribbon}b as $\omega = 2.83$~eV. The emission of the magnetic dipole source placed in the plasmonic lattice will be modified due to the surrounding environment. This is characterised by the Purcell factor $P_F$, the ratio of emitted power of a magnetic dipole to the emitted power in free space $P_M$ \cite{baranov2017modifying},
\begin{equation}
    P_M = \frac{\mu_0}{4\pi}\frac{\omega^4|\textbf{m}|^2}{3c^3},
\end{equation}

where $\mu_0$ is the vacuum permeability and $\textbf{m}$ is the magnetic dipole moment. The Purcell factor as a function of source position is shown in \autoref{fig:directionality}b (dotted orange line). The enhancement is greatest when the source is close to the metallic nanoparticles in the lattice, but we note it is generally modest due to the distance between the source and nanoparticles. 

To characterise the directionality of the energy flow along the edge, we integrate the Poynting vector through a plane perpendicular to the interface and metasurface, and we plot this as solid lines in \autoref{fig:directionality}b; with purple and green corresponding to the fraction of flow to the right ($P_R$) and left ($P_L$), respectively. Regions in which the coupled dipole approximation does not hold are shaded. Starting at $\textbf{r}_1$, and looking at the power flow for the right circular polarisation, the flow of energy is predominantly to the right ($P_R$, purple line) as expected from the polarisation of the source, which couples to the right-propagating pseudospin mode. Importantly there is still a fraction of energy travelling to the left ($P_L$, green line) which demonstrates the existence of pseudospin mixing. 
As the source is moved towards $\textbf{r}_2$ the flow is completely to the right before quickly flipping to the opposite direction. From $\textbf{r}_2$ to $\textbf{r}_3$ the energy flow again changes sign as the source moves from an area with negative spin angular momentum to an area with positive spin angular momentum. 
The propagation direction of the excited edge state can then be predicted from the interplay between the source polarisation and the local handedness of the polarisation of the mode, given by the spin angular momentum in real space, rather than by the pseudospin. More details on the directionality of the modes excited by dipole sources are given in the Supporting Information, Figure S1).
We emphasise that we have used a right hand polarised source throughout, which suggests one should always expect energy flow in the right direction. 

In \autoref{fig:directionality}c-d we show the stark difference in directionality along the edge depending on the position of the source (magenta star). We consider the same right-circularly polarized source placed at two different positions. In \autoref{fig:directionality}c we choose the optimal placement for excitation in the expected direction, and in \autoref{fig:directionality}d the least optimal placement. All bosonic systems with these lattice symmetries, such as the photonic crystal, will also possess these position dependent directional modes when excited with a circularly polarised source \cite{oh2018chiral}. In the left and right panels, we show the electric field intensity, $|\textbf{E}|^2$, at plane cut perpendicular to the metasurface in the left and right directions. We note that the edge state is not only confined to the edge in the plane of the metasurface but also out of the plane at subwavelength scales. Finally, we stress that in this discussion we have considered sources placed in the plane of the metasurface. For sources above the nanoparticles, the excitation of directional edge modes will be determined by, first, the value of the Purcell factor, and, second, the interplay between the source polarization and the distribution of spin angular momentum. (Plots of the spin angular momentum in planes above the metasurface are given in the Supporting Information, Figure S2.)

\section*{Retardation and Radiative Effects}\label{sec:retardation}
We finally discuss in detail the effect of retardation and radiation in the pseudospin edge states. While so far we have considered a very subwavelength period and nanoparticle size, we have already seen the effect of retarded interactions which cause the band structure to be altered with respect to the QSA, in particular close to the light line, and the radiative broadening and redshifting of resonances. Radiative effects become more apparent for larger nanoparticles, with very large broadenings and shifts as in the single particle extinction cross section shown in \autoref{fig:lattice_layout}d; effects which are not captured in the QSA. On the other hand, it is important to note that retardation can have striking consequences not only on bulk band structures but also on the properties of edge states. It has been shown for example how in the 1D plasmonic SSH model, retardation can result in the breakdown of bulk boundary correspondence and the disappearance of edge states \cite{pocock2019bulk}. To investigate the effects of retardation in the 2D system studied here, we excite an interface between expanded and shrunken regions in a ribbon with a plane wave at a finite wavevector above the light line, $k_x = 0.15\pi/a_0$, and calculate the extinction cross section for increasing lattice constants $a_0$.

\begin{figure}
    \includegraphics[width=0.5\textwidth]{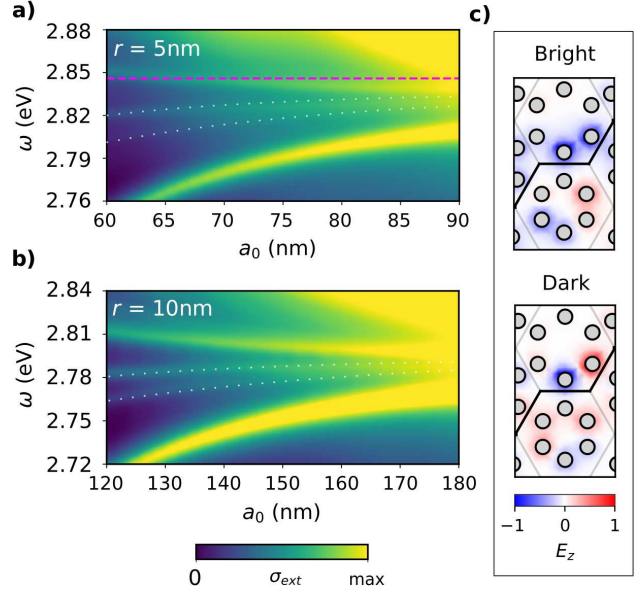}
    \caption{Far-field excitation of pseudospin edge states showing radiative and retardation effects. Extinction cross section for ribbon system under plane wave excitation at $k_x = 0.15\pi/a$ for increasing lattice constants $a_0$ with Drude losses $\gamma = 0.01$~eV. The edge states are highlighted (white dotted lines), a) For particle radius $r=5$~nm and height $h=20$~nm. The surface plasmon frequency $\omega_{sp}$ is shown as the purple dotted line, b) For particle radius $r=10$~nm and height $h=40$~nm. The bands are shifted to lower frequencies, far below $\omega_{sp}$, c) Normalised out-of-plane electric field, in the metasurface plane, for the upper and lower energy edge states for the parameters in a). The upper mode is bright as it has a bonding symmetry along the edge, allowing it to be excited by a plane wave, while the lower mode is dark.}
    \label{fig:retardation}
\end{figure}

In \autoref{fig:retardation}a, we show the extinction cross section for a metasurface with an interface between the two phases for silver nanoparticles with radius $5$~nm and height $20$~nm. We let the Drude losses $\gamma = 0.01$~eV, as in \autoref{fig:ribbon}b, and highlight the edge states with white dots for visibility. As in the infinite lattice, in the quasistatic and nearest neighbour approximations, the edge states are expected to be symmetrical about the plasma frequency $\omega_{sp}$ (dotted purple line) \cite{honari2019topological}. In contrast, when radiative effects are taken into account, there are substantial shifts in the edge state frequencies. Initially, for $a_0 = 60$~nm the edge states are well separated from the bulk but as the lattice constant increases up to $90$~nm the bulk modes close up and the edge states are lost. We calculate the full width at half maximum (FWHM) of the highest energy edge state at $a_0 = 60$~nm where FWHM $= 0.0059$~eV. As an aside, this edge state has a larger cross section as along the interface the nanoparticles form a bonding-like state (\autoref{fig:retardation}c, top) whereas the lower energy edge state has an anti-bonding distribution which leads to lower cross section (\autoref{fig:retardation}c, bottom). In \autoref{fig:retardation}b we show the cross section for nanoparticles with radius $10$~nm and height $40$~nm, with lattice constant varying from $120$ to $180$~nm. Here, the edge states become significantly redshifted far below $\omega_{sp}$. Again we measure the FWHM, at $a_0 = 120$~nm, FWHM $ = 0.0077$~eV demonstrating the broadening of the mode for larger nanoparticles. We note that these widths are smaller than the Drude losses since the lattice structure modifies the optical response by increasing the quality factor \cite{giannini2011plasmonic, kravets2018plasmonic}.

\section*{Conclusion}

In this work we have presented a study of spin dependent edge states in a plasmonic metasurface. These states rely on lattice symmetries and are similar to the quantum spin Hall effect in topological insulators. By going beyond the quasistatic approximation and including retardation and radiative effects, we model the plasmonic system appropriately and show how the long range interactions result in a more complex band structure than the QSA. The bands involved in the band inversion are not greatly affected by retardation, and they provide a signature for far-field measurements. 
We note that the ordering in energy of the modes of the plasmonic metasurface are opposite to what is seen in photonic crystals \cite{wu2015scheme}, with the dipolar modes lying at higher energies than the quadrupolar modes for the non-inverted bands. 

Importantly, we determine the spin angular momentum of the edge modes, and show that this is the quantity that characterises the directionality of the modes. Remarkably, this leads to a more complex behaviour which can be opposite to what would be expected from a direct analogy with the QSH effect. We emphasise that these conclusions are valid not only for the plasmonic metasurface but for any bosonic system with this lattice \cite{oh2018chiral}. We probe the spin angular momentum by looking at the excitation of the edge states in the near-field by a magnetic dipole source. By varying the position of the source, we highlight how the location is essential in exciting a purely unidirectional state. In agreement with the spin angular momentum characterisation of the edge in the ribbon we show that in some positions a source will excite a mode in completely the opposite way to the expected direction. 

Finally, we have also considered the optical response of the plasmonic metasurface and the edge modes under far-field excitation, showing the effect of radiative and retardation effects for the larger radius nanoparticles and larger lattice constants. Although the edge states persist for larger nanoparticles, we observe a radiative broadening and shift in frequency which is naturally not captured in a quasistatic approximation. For increasing lattice constants, the bulk modes close up and eventually the edge states are indistinguishable, which provides a parameter regime for which these edge states could be experimentally observed.

\section*{Methods}

\subsection*{Lattice sums}\label{sec:methods_sums}

The summations in the interaction matrix in \autoref{eqn:hamiltonian} are conditionally convergent. Firstly, we will consider sums including the origin term and split these into long range and short/medium range terms,
\begin{align}
    S_{\text{incl}} = \sum_\textbf{R} e^{i\textbf{k}\cdot\textbf{R}} \hat{\textbf{G}}(\textbf{r}, \omega) = S_L + S_{SM},
\end{align}
where $S_L = k^2\sum_\textbf{R} e^{i\textbf{k}\cdot\textbf{R}}e^{ikd}\frac{1}{d}$ and $S_{SM} = k^2\sum_\textbf{R} e^{i\textbf{k}\cdot\textbf{R}}e^{ikd}\left(\frac{ik}{d^2} - \frac{1}{d^3}\right)$. The slowly converging $S_L$ term is handled by using Ewald's method \cite{linton2010lattice}. This splits the real space sum into two and then takes the Fourier transform of one part using Poisson's summation, resulting in a sum over the reciprocal lattice. The sum is optimised with a Ewald parameter to ensure the real space and reciprocal space sum converge within approximately the same number of lattice constants. The $S_{SM}$ term converges rapidly above the light line. Outside of this region the first few terms within a radius $R_{min}$ are added and then the rest of the sum is calculated numerically by approximating the summation as an integral \cite{zhen2008collective},
\begin{align}
    S_{SM} \approx k^2\sum_{\textbf{R}=0}^{\textbf{R}_{min}} e^{i\textbf{k}\cdot\textbf{R}}e^{ikd}\left(\frac{ik}{d^2} - \frac{1}{d^3}\right) + k^2 \int_{R_{min}}^{\infty}e^{i\textbf{k}\cdot\textbf{R}}e^{ikd}\left(\frac{ik}{d^2} - \frac{1}{d^3}\right).
\end{align}

For sums excluding the origin, 
\begin{align}
    S_{\text{excl}} = \sum_{\textbf{R}\neq0} e^{i\textbf{k}\cdot\textbf{R}} \hat{\textbf{G}}(\textbf{R}, \omega),
\end{align}
the integral method from Ref. \onlinecite{zhen2008collective} is used. For ribbons which are infinite in only one direction, the summations converge easier and so no techniques are used to speed up convergence \cite{wang2016existence}. 

In the quasistatic approximation, we only consider the short range $1/d^3$ term which converges quickly when including all neighbours, in both the infinite lattice and semi-infinite ribbon. We note that this method of including all neighbours results in a kink at $\Gamma$ in one of the modes of the infinite lattice which is due to the group velocity necessarily being zero at the centre of the BZ \cite{zhen2008collective}.

\subsection*{Linearised Green's Function}

In \autoref{fig:ribbon}, we linearise the Green's function to investigate the spin angular momentum semi analytically. For the retarded Green's function and radiative polarisabilities the eigenvalue problem is non-linear and non-Hermitian,
\begin{align}\label{eqn:supp_nonlinear_eigval}
    \left(\hat{\textbf{H}}(\textbf{k}, \omega) - \frac{1}{\alpha(\omega)}\hat{\textbf{I}}\right)\cdot\textbf{p}=0.
\end{align}

To avoid the computational complexity of searching for complex $\omega$ solutions we linearise the Green's function by making the approximation $\omega = \omega_{sp}$, the surface plasmon frequency. In spherical nanoparticles, $\omega_{sp} = \omega/\sqrt{\epsilon_\infty + 2}$ and for spheroids, $\omega_{sp} = \omega/\sqrt{\epsilon_\infty - 1 + 1/L}$. For out of plane modes in the system we consider, $\omega_{sp} = \omega/3.12$. This is valid for the size of nanoparticles considered here since $\omega$ varies faster in the polarisabilty term than in the Green's function. As particle size increases the approximation becomes less valid close to the light line. We can then rewrite \autoref{eqn:supp_nonlinear_eigval} as, 
\begin{align}\label{eqn:supp_nonlinear_eigval}
    \left(\hat{\textbf{H}}(\textbf{k}, \omega_{sp}) - \frac{1}{\alpha(\omega)}\hat{\textbf{I}}\right)\cdot\textbf{p}=0,
\end{align}
for which eigenvalues $\lambda$ and eigenvectors $\textbf{p}$ are found at each point in the BZ, $\textbf{k}$, and band structures are calculated by rearranging $\lambda = 1/\alpha(\omega)$ to find $\omega$, for the static polarisability.

\subsection*{Extinction Cross Section}

To calculate the extinction cross section, $\sigma_{\text{ext}}$, we use the following system of equations for a non-zero external field,
\begin{align}\label{eqn:methods_non_zero_field}
    \left(\hat{\textbf{H}}(\textbf{k}, \omega) - \frac{1}{\alpha(\omega)}\hat{\textbf{I}}\right)\cdot\textbf{p} = \textbf{E}_{\text{inc}}.
\end{align}
From Maxwell's equations, we have $k_\parallel E_\parallel + k_z E_z = 0$ and we assume $E_\parallel = 1$. Rearranging, we can then write,
\begin{align}
    E_z = \frac{-k_\parallel}{k_z} = \frac{-k_\parallel}{k_0^2 - k_\parallel^2}.
\end{align}
The total incident field on each particle includes an additional phase due to the position within the unit cell $\textbf{d}$, $\textbf{E}_{inc} = \textbf{E}\exp{(i\textbf{k}\cdot\textbf{d})}$. After calculating dipole moments using \autoref{eqn:methods_non_zero_field}, the extinction cross section is given by the optical theorem,
\begin{align}
    \sigma_{\text{ext}} = \frac{4\pi k \sum_i \text{Im}(\textbf{p}_i \cdot \textbf{E}_{\text{inc}}^*)}{|\textbf{E}|^2}.
\end{align}

\subsection*{Spectral Function}

The spectral function method relies on an effective polarisability formulation of the system \cite{zhen2008collective, wang2016existence}. 
We rewrite the system of equations for a non-zero external field as $\textbf{p}=\alpha_{\text{eff}}\textbf{E}$. The effective polarisability $\alpha_{\text{eff}} = 1/\lambda$ for eigenvalues of $\textbf{M}$, $\lambda$. The spectral function is analogous to the extinction cross section but rather than describing the system when excited by a well defined external field instead characterises all modes in the system in the retarded, radiative regime, regardless of whether they are bright or dark modes. This corresponds to the forced oscillation of each mode of the lattice at some driving frequency $\omega$ and Bloch wavevector $\textbf{k}$. The spectral function is defined,

\begin{align}
    \sigma_{\text{spectral}} = 4\pi k\sum_i \text{Im}(\alpha_{\text{eff}}^{(i)}),
\end{align}

where the sum is over the number of elements in the unit cell in the infinite lattice or the super cell in the semi-infinite ribbon. Peaks in the spectral function will correspond to the real part of the band structures from the linearised Green's function.

\begin{acknowledgement}

We acknowledge fruitful discussions with A. Garc\'ia-Etxarri and Mehul P. Makwana. M.P., R.V.C. and P.A.H. acknowledge funding from the Leverhulme Trust. P.A.H. also acknowledges funding from Funda\c c\~ao para a Ci\^encia e a Tecnologia and Instituto de Telecomunica\c c\~oes under project CEECIND/03866/2017. S.A.M. and R.V.C. acknowledge funding from EPSRC Programme Grant ``Mathematical Fundamentals of Metamaterials'' (EP/L024926/1). S.A.M. additionally acknowledges the Lee-Lucas Chair in Physics. We thank Jude Marcella for the TOC figure.

\end{acknowledgement}

\begin{suppinfo}
Position dependence on directionality of the edge states, with left and right polarised magnetic dipole sources. Spin angular momentum calculated at planes above the metasurface.
This material is available free of charge on the \href{http://pubs.acs.org}{ACS Publications} website.
\end{suppinfo}

\bibliography{bibliography.bib}

\end{document}


\maketitle

{\centering Comprised of:\\ 2 pages,\\ 2 figures,\\ 0 tables\\
}
\pagenumbering{gobble}

\clearpage

\pagenumbering{arabic}

\renewcommand{\thepage}{S\arabic{page}} 
\renewcommand{\thesection}{S\arabic{section}}  
\renewcommand{\thetable}{S\arabic{table}}  
\renewcommand{\thefigure}{S\arabic{figure}}

\section*{Position dependence of magnetic dipole sources on directionality}
\begin{figure}[h]
    \centering
    \includegraphics[width=\textwidth]{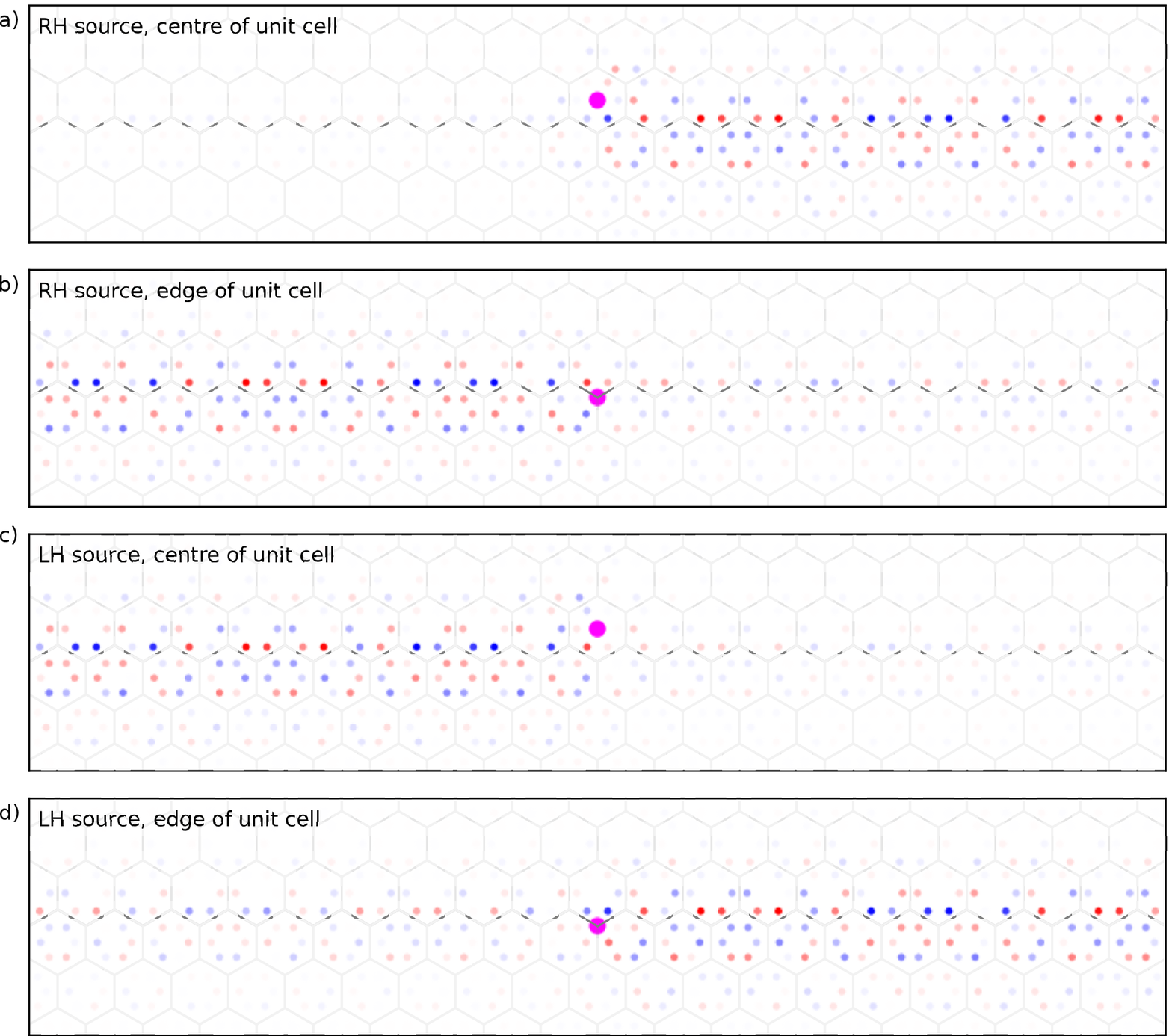}
    \caption{The position dependent directionality of edge states. We excite a mode with a magnetic dipole source with the same properties as in the main text at the centre of the unit cell next to the edge and on the edge itself. a) When the source is at the centre of the cell, directionality is as we expect from a right handed source, b) For a source at the edge directionality is opposite, c) Expected directionality from a left handed source placed at the centre of the cell, d)  Directionality is opposite for a left handed source placed at the edge.}
    \label{fig:supp_directionality}
\end{figure}

\section*{Spin angular momentum above the plane of the metasurface}

\begin{figure}
    \centering
    \includegraphics[width=\textwidth]{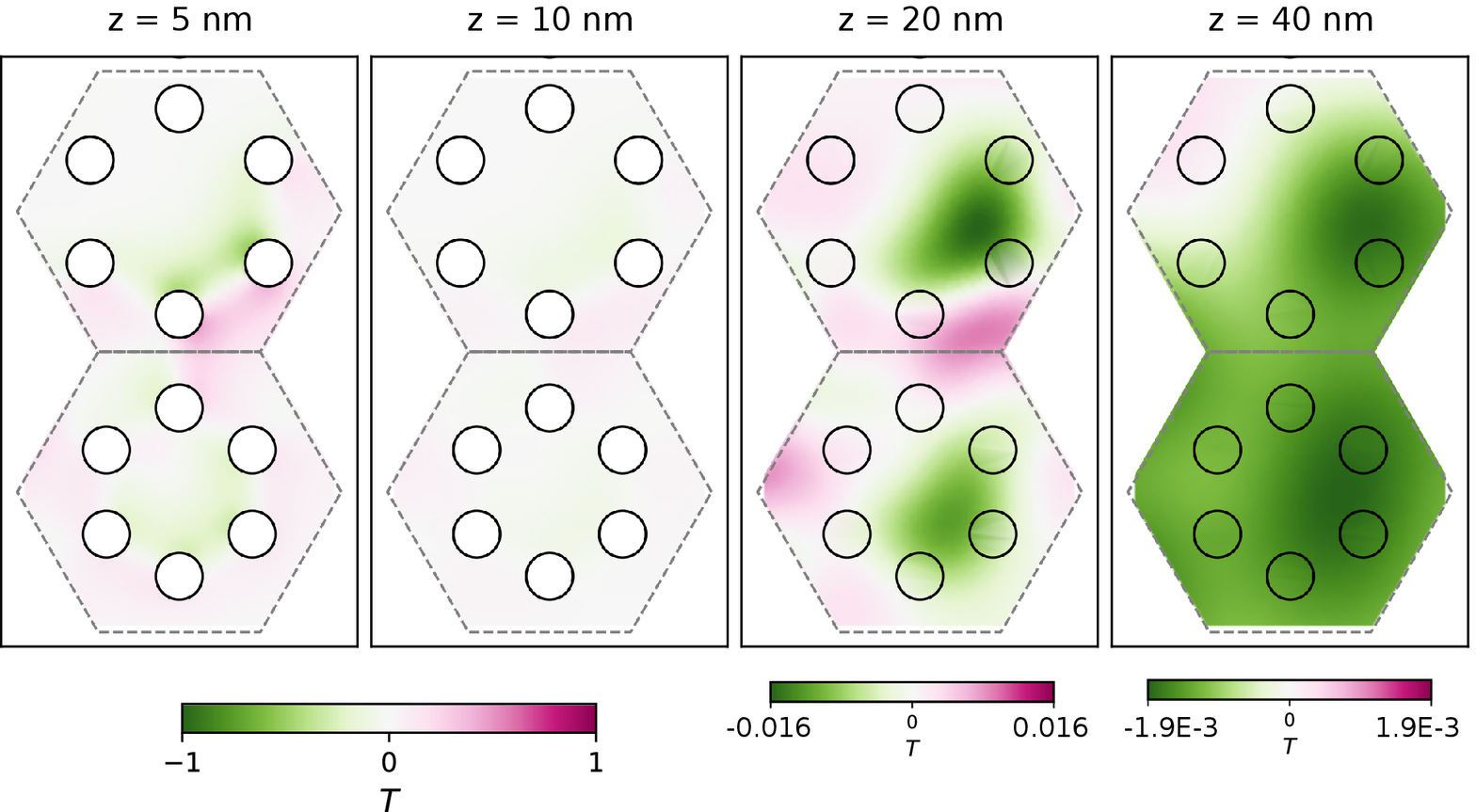}
    \caption{Spin angular momentum $T_z$ calculated at planes above the metasurface, for mode $s_2$ in the main text. For $z=5$~nm and $10$~nm we normalise to the maximum value at $z=0$. The inhomogeneity pattern of $T_z$ is still present for $z=5$~nm, but soon becomes less visible for $z=10$~nm. For larger $z$, $T_z$ becomes more homogeneous but the magnitude is significantly smaller than in the plane of the metasurface.}
    \label{fig:supp_pseudo_planes}
\end{figure}